\documentclass[aps,pra,superscriptaddress,twocolumn,footinbib,showpacs,floatfix]{revtex4-1}

\usepackage{graphicx}
\usepackage{dcolumn}
\usepackage{bm}
\usepackage{color}
\usepackage[colorlinks=true,urlcolor=blue,citecolor=blue, linkcolor = blue]{hyperref}
\usepackage{amssymb,ulem,amsmath}
\usepackage[percent]{overpic}
\def\be{\begin{equation}}
\def\ee{\end{equation}}
\def\ba{\begin{array}{lll}}
\def\ea{\end{array}}
\def\ber{\begin{eqnarray}}
\def\eer{\end{eqnarray}}

\newcommand{\qql}{\textquotedblleft}
\newcommand{\qqr}{\textquotedblright}

\begin{document}
\title{Superconducting memristors}
\author{Sebastiano Peotta}
\email{speotta@physics.ucsd.edu}
\author{Massimiliano Di Ventra}
\email{diventra@physics.ucsd.edu}
\affiliation{Department of Physics, University of California, San Diego, La Jolla, CA 92093, USA}

\begin{abstract}
In his original work Josephson predicted that a phase-dependent conductance should be present in superconducting tunnel junctions, an effect difficult to detect, mainly because it is hard to single it out from the usual non-dissipative Josephson current.
We propose a solution for this problem that consists in using different superconducting materials to realize the two junctions of a superconducting interferometer. According to the Ambegaokar-Baratoff relation the two junctions have different conductances if the critical currents are equal, thus the Josephson current can be suppressed by fixing the magnetic flux in the loop at half of a flux quantum without cancelling the phase-dependent conductance. Our proposal can be used to study the phase-dependent conductance, an effect present in principle in all superconducting weak links. From the standpoint of nonlinear circuit theory such a device is in fact an ideal memristor with possible applications to memories and neuromorphic computing in the framework of ultrafast and low energy consumption superconducting digital circuits.
\end{abstract}
%
%
\maketitle

\section{Introduction}\label{sec:introduction}

\begin{figure}[t]
\includegraphics[scale = 1.1]{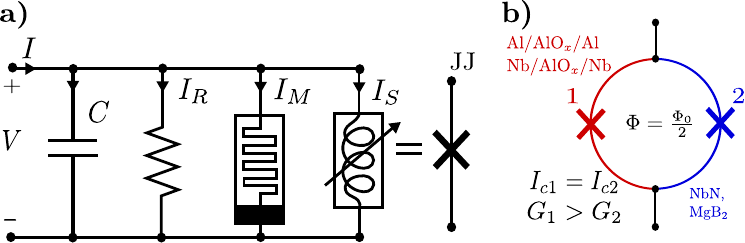}
\caption{\label{fig1} \textbf{a)} Equivalent circuit of a Josephson junction corresponding to Eq.~(\ref{eq:josephson1}) with, in order, capacitance $C$, quasiparticle dissipative current $I_R = GV$, phase-dependent dissipative current $I_M = \varepsilon G\cos\gamma\, V$ and Josephson current $I_S = I_c\sin\gamma$. The phase-dependent dissipative current is represented with the circuital symbol of a memristor. \textbf{b)} A conductance asymmetric two-junction interferometer (CA-SQUID) working as a memristor. The two junctions have ideally the same critical currents $I_{c,1} = I_{c,2}$, but unequal conductances $G_1 \neq G_2$ since the electrodes are made of different superconducting materials (denoted by different colors). The magnetic flux $\Phi = \Phi_0/2$ suppresses the total critical current of the interferometer ($\Phi_0= h/(2e)$ is the flux quantum), but not the phase-dependent dissipative current due to the conductance asymmetry.}
\end{figure}

The basic circuit element of superconducting electronics is the Josephson junction (JJ), a tunnel barrier between two superconductors~\cite{Josephson:1962,Barone_book}, characterized by a non-dissipative current $I_S = I_c\sin \gamma(t)$ (Josephson current), where $I_c$ is the critical current and $\gamma(t)$ is the gauge-invariant phase difference between the order parameters of the two superconducting electrodes~\cite{Tinkham_book}. Alongside this term, Josephson~\cite{Josephson:1962} predicted an additional phase-dependent dissipative current $I_M = G(\gamma)V$, with $V$ the voltage drop across the junction (see Fig.~\ref{fig1}\textbf{a}). The phase-dependent conductance (PDC) $G(\gamma) \propto \cos\gamma$ can be interpreted either as an interference effect between quasiparticle and Cooper pair currents~\cite{Tinkham_book,Josephson:1962} or, alternatively, as a consequence of the retarded phase-current response~\cite{Harris:1974,Harris:1975,Harris:1976,Zorin:1979}.

The measured value of the PDC in tunnel junctions~\cite{pfl,smp}, point contacts~\cite{vd,rd} and weak links~\cite{nw,fpt,Likharev:1979} can not be explained by BCS theory. Several effects that can account for the discrepancy have been discussed in Ref.~\cite{Zorin:1979}, but little is known about the role and/or use of this effect in actual devices and the subject is still regarded as controversial.
Recently, the PDC has been discussed theoretically in Refs.~\cite{Catelani:2011a,Leppakangas:2011}
and studied in an experiment on fluxonium qubits~\cite{Pop:2014} aimed at understanding quasiparticle-induced decoherence in superconducting qubits~\cite{Moji:1999,Lutchyn:2007,Makhlin:2001,Martinis:2009,Lenander:2011,Catelani:2011b,Nori:2005,Wilhelm:2008,Devoret:2013}, 
but an easy way to isolate this term from the Josephson current has not been suggested so far.

In this Article we propose to isolate the PDC from the non-dissipative current $I_S$ using a two-junction interferometer as shown in Fig.~\ref{fig1}\textbf{b}, which can be tuned to have a vanishing total critical current but a finite residual PDC since the two junctions are made of \textit{different superconducting materials}. Such a \textit{conductance-asymmetric} SQUID (CA-SQUID) has never been used so far to study the PDC.

A further motivation of our proposal is  that a circuit element defined by
\begin{gather}
I(t) = G(\phi(t))V(t)\quad \text{and} \quad \frac{d \phi(t)}{dt} = V(t),\label{eq:memristor1}
\end{gather}
with $G(\phi) >0$, is called an ideal {\it memristor}~\cite{chua3,chua1,Pershin:2011}, a dissipative element whose resistance is a function of an internal memory degree of freedom, the flux linkage $\phi = \int dt\,V$ in the case of Eq.~(\ref{eq:memristor1}). The research in the field of memristive devices has been flourishing since the unambiguous identification of memory behavior in TiO$_2$ cross-point switches~\cite{strukov}, which have promising applications~\cite{Pershin:2011,Likharev:2011}. In a specific regime to be discussed below a CA-SQUID is described by Eq.~(\ref{eq:memristor1}) with $\phi(t) \propto \gamma(t)$ and is a new --- superconducting --- implementation of an ideal memristor.
The aim of our work is then to clarify the conditions for a two-junction interferometer to be described by Eq.~(\ref{eq:memristor1}) and to pinpoint novel manifestations of the PDC inspired by the theory of memristive circuit elements and ignored so far in the context of JJs.

The structure of this Article is as follows.
In Sec.~\ref{sec:pdc} we introduce the concept of PDC from the microscopic theory as a phenomenon generally present in Josephson tunnel junctions and we justify the JJ models used in the remainder of the work. The idea to isolate the PDC of a suitably engineered two-junction interferometer, thereby realizing a superconducting memristor, is explained in Sec.~\ref{sec:ideal_memristor}. In Sec.~\ref{sec:experimental_consideration} the conditions that need to be satisfied in a realistic implementation are elucidated. The subject of Sec.~\ref{sec:pinched_hysteresis_loop} are pinched hysteresis loops in the $I-V$ plane, the most distinctive feature of memristors, that in the present case have unique properties. In Sec.~\ref{sec:nondestructive_readout} we propose an alternative way to probe the PDC using single-flux-quantum voltage pulses. It is shown how the internal state of a superconducting memristor can be extracted without changing its value. This non-destructive readout protocol may be of interest for superconducting memories compatible with Rapid Single Flux Quantum (RSFQ) superconducting circuits~\cite{Likharev:1991,Bunyk:2001,Likharev:2012}. In Sec.~\ref{sec:fluctuations} we analyse the form of the current noise for a superconducting memristor and derive the corresponding drift-diffusion equation for the probability distribution of the phase in the limit of high damping. Finally we summarize our results and discuss possible future developments in Sec.~\ref{sec:conclusions}.

\begin{figure*}[t]
\includegraphics{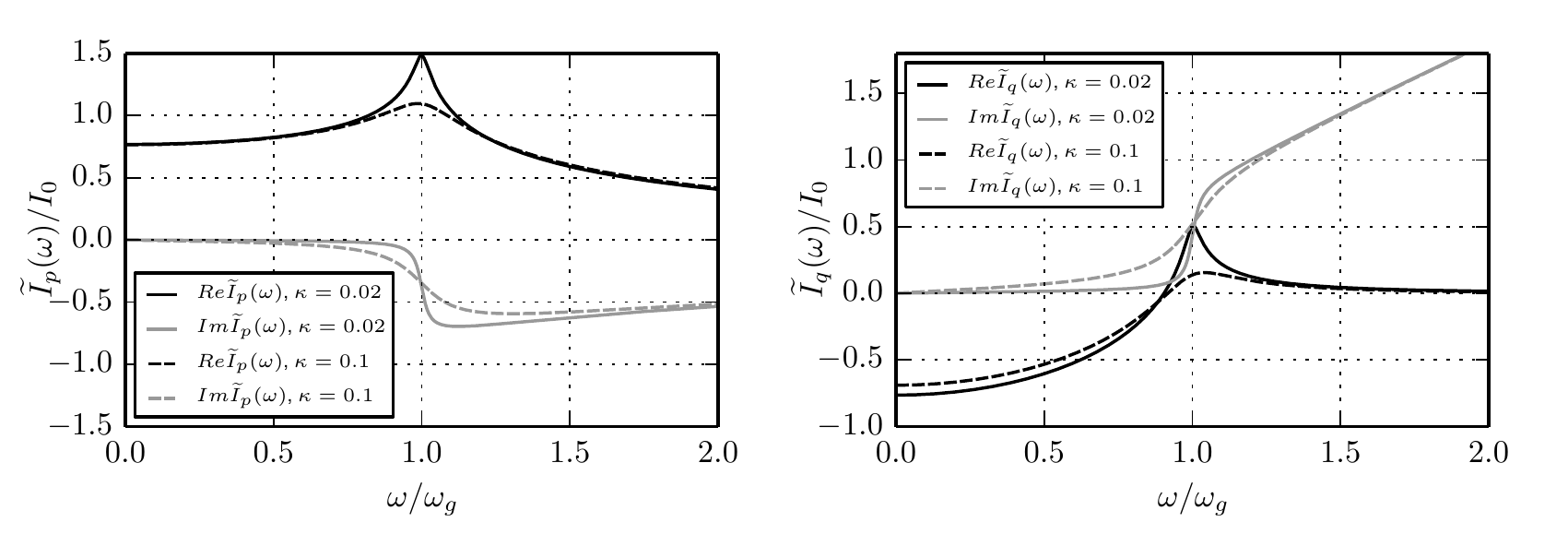}
\caption{\label{fig:current_fourier} Fourier transforms $\widetilde{I}_p(\omega)$ and $\widetilde{I}_q(\omega)$ of the phase-current response functions $I_p(t)$ and $I_q(t)$ given by Eq.~(\ref{eq:Ip}) and~(\ref{eq:Iq}). Frequency is in units of the gap frequency $\omega_g = 2\Delta/\hbar$ and the current in units of $I_0 = G_N V_g = G_N (2\Delta/e)$. Increasing $\kappa$ has the effect of increasing the normal subgap current $Im \widetilde{I}_q(\omega_J/2)$ and the phase-dependent dissipative current $Im\widetilde{I}_{p}(\omega_J/2)\cos\omega_J t$ for $\omega_J/2 < \omega_g$, while their ratio $\varepsilon$ does not vary much ($\omega_J = 2eV_0/\hbar$ with $V_0$ the constant bias voltage). Moreover the peak in the Josephson current $Re \widetilde{I}_p(\omega_J/2)\sin\omega_J t$ (Riedel peak) is smoothed out.}
\end{figure*}

\section{Phase-dependent conductance}\label{sec:pdc}
Our idea for realizing an ideal memristor rests on the use of the PDC of a JJ that, from a theoretical standpoint, is as fundamental as the DC and AC Josephson effects, but has received much less attention in the past possibly due to its intrinsically dissipative and AC character as opposed to the non-dissipative Josephson effect. In the following we provide a short introduction to this effect and at the same time justify the simple JJ model that we use in the rest of the work.

The dynamics of a generic low-transparency JJ are well described by second order perturbation theory in the tunneling matrix elements resulting in the Tunnel Junction Microscopic (TJM) model~\cite{Likharev_book} where the total current $I = I_\text{pair}+I_\text{qp}$ is the sum of the pair current $I_\text{pair}$ and quasiparticle current $I_\text{qp}$, given by
\begin{gather}\label{eq:tjm1}
I_\text{qp}(t) = \int_{-\infty}^{t} dt'\, I_q(t-t')\sin\left(\frac{\gamma(t)-\gamma(t')}{2}\right)\,,\\
I_\text{pair}(t) = \int_{-\infty}^{t} dt'\, I_p(t-t')\sin\left(\frac{\gamma(t)+\gamma(t')}{2}\right)\,.
\label{eq:tjm2}
\end{gather}
A time-dependent phase corresponds to a finite voltage drop across the junction~\cite{Josephson:1962}
\begin{equation}\label{eq:josephson2}
\frac{d\gamma(t)}{dt} = \frac{2e}{\hbar}V(t) = \frac{2\pi}{\Phi_0}V(t)\,,
\end{equation}
with $\Phi_0 = h/(2e)$ the flux quantum.
For constant phase $\gamma(t)=\gamma_0$ Eq.~(\ref{eq:tjm2}) gives the Josephson current $\propto \sin\gamma_0$ while Eq.~(\ref{eq:tjm1}) is nonzero only for a time-dependent phase.
The temperature and the material properties of the superconducting electrodes and barrier layer enter only in the nonlinear phase-current response functions $I_p(t)$ and $I_q(t)$. We consider for definiteness the result obtained from BCS theory at zero temperature, which can be stated in closed form~\cite{Harris:1976}, with an additional phenomenological exponential factor:
\begin{gather}
I_p(t) = -\frac{2I_c}{\tau_g}J_0\left(\frac{t}{\tau_g}\right)Y_0\left(\frac{t}{\tau_g}\right)\exp\left(-\frac{t}{\tau_r}\right)\,,\label{eq:Ip}\\
I_q(t) = \frac{2I_c}{\tau_g}J_1\left(\frac{t}{\tau_g}\right)Y_1\left(\frac{t}{\tau_g}\right)\exp\left(-\frac{t}{\tau_r}\right)-\frac{\hbar G_N}{e}\delta'(t)\,,
\label{eq:Iq}
\end{gather}
where $J_n\,,Y_n$ are the Bessel functions of the first and second type, respectively, $\delta'(t)$ is the derivative of the delta function, and $\tau_g = \hbar/\Delta$ with $\Delta$ the superconducting gap of the electrodes. $I_c$ is the critical current and $G_N$ the normal junction conductance. The exponential factor $\exp(-\tau/\tau_r)$ is not a result of the BCS theory but accounts for the experimentally observed broadening of the Riedel peak~\cite{Likharev_book} (see Fig.~\ref{fig:current_fourier}) and introduces a finite retardation time $\tau_r$ for the otherwise slowly decaying functions $J_n(t)Y_n(t) \sim t^{-1}\cos 2t$. The peak broadening is well captured by $\kappa = 0.03\div 0.1$, with $\kappa= \tau_g/(2\tau_r)$, in most tunnel junctions~\cite{Likharev_book}. The Fourier transforms of Eqs.~(\ref{eq:Ip})-(\ref{eq:Iq})  are shown in Fig.~\ref{fig:current_fourier} for two values of $\kappa$  and correspond to the various components of the current under a DC voltage bias~\cite{Likharev_book}.

The considerations that follows are largely independent of the specific form of the response functions $I_{p,q}(t)$ which are only required to decay sufficiently fast for $t\gtrsim \tau_r$, a physically reasonable assumption. However the expressions provided in Eqs.~(\ref{eq:Ip})-(\ref{eq:Iq}) allow for a simple analytic result (details are in Appendix~\ref{appendix:analytic_pdc}).
We assume the phase $\gamma$ to be slowly varying $|\dot\gamma| \tau_r = 2e|V|\tau_r/\hbar\ll 1$, thus $\gamma(t)-\gamma(t') \approx \frac{2eV}{\hbar}(t-t')$ for $|t-t'| \lesssim \tau_r$. Using this approximation and the identity
\begin{equation}\begin{split}
\sin\tfrac{1}{2}(\gamma(t)+\gamma(t')) &= \sin\gamma(t)
\cos\tfrac{1}{2}(\gamma(t)-\gamma(t')) \\ &-\cos\gamma(t)
\sin\tfrac{1}{2}(\gamma(t)-\gamma(t'))\,,
\end{split}\end{equation}
in Eqs.~(\ref{eq:tjm1})-(\ref{eq:tjm2}),  gives for the total junction current
\begin{equation}\label{eq:jop}
\begin{split}
I(t) &\approx -2I_cA_{0,0}(\kappa)\sin\gamma(t) \\&+G_N\left[1+\pi A_{1,1}(\kappa) + \pi A_{0,1}(\kappa)\cos\gamma(t)\right]V(t)\,,
\end{split}
\end{equation}
where the functions $A_{n,m}(\kappa)$ with $n,m$ integer numbers are defined by
\begin{equation}\label{eq:A_nm}
A_{n,m}(\kappa) = \int_0^{+\infty}dx\,x^mJ_n(x)Y_n(x)e^{-2\kappa x}\,,
\end{equation}
and can be evaluated in terms of elliptic integrals (see Appendix~\ref{appendix:analytic_pdc}). We neglect the correction to the critical current provided by the factor $-2A_{0,0}(\kappa) \sim 1$ and define $G_L(\kappa) = G_N[1+\pi A_{1,1}(\kappa)] < G_N$ as the leakage conductance, i.e. the conductance for $V < V_g = 2\Delta/e$, and the coefficient $\varepsilon(\kappa) = \pi A_{0,1}(\kappa)/[1+\pi A_{1,1}(\kappa)]$, which satisfies the condition $|\varepsilon| <1$.
Therefore in the low frequency (low voltage) regime --- namely for $V \lesssim V_g$ since the retardation time $\tau_r$ is of the same order of $\tau_g = \hbar/\Delta$ --- the TJM model is well approximated by the Resistively Shunted Junction (RSJ) model~\cite{Likharev_book} whose equivalent circuit is shown in Fig.~\ref{fig1}\textbf{a} and reads
\begin{equation}
I = C\frac{dV}{dt} + G_L(1+\varepsilon\cos\gamma)V + I_c\sin \gamma +I_F(t)\,.\label{eq:josephson1}
\end{equation}
In addition to the quasiparticle~(\ref{eq:tjm1}) and pair~(\ref{eq:tjm2}) currents, we have added the displacement current $C\frac{dV}{dt}$ and the current fluctuations $I_F(t)$.
A finite capacitance $C$ must be introduced in order to account for the geometrical capacitance of the electrodes and higher order terms in the previous expansion of Eqs.~(\ref{eq:tjm1}) and~(\ref{eq:tjm2}). The latter effect is small and negligible with respect to the PDC. The fluctuating current term $I_F(t)$ is discussed in Sec.~\ref{sec:fluctuations}.

The ratio $\varepsilon$ between the PDC and the leakage conductance has been investigated in a number of experiments on tunnel junctions~\cite{pfl,smp}, point contacts~\cite{vd,rd} and weak links~\cite{nw,fpt} finding consistently $\varepsilon\sim -1$ at low temperature, while BCS theory predicts $\varepsilon > 0$~\cite{Likharev_book,Harris:1974}. As shown in Appendix~\ref{appendix:analytic_pdc} the exponential regularization  gives $\varepsilon \approx -1/3$, with the sign in agreement with experimental results.

We emphasise that a finite PDC is ultimately due to the fact that the phase $\gamma(t')$ at an earlier time $t'<t$ enters the expression~(\ref{eq:tjm1}) for the pair current $I_\text{pair}(t)$ and it is not a consequence of specific properties of the response functions.
Indeed an hand-waiving argument to deduce the existence of the \qql$\cos\gamma$\qqr term is to include a retardation directly in the first Josephson equation $I_S=I_c\sin\gamma(t-\tau_r) \approx I_c\left[\sin\gamma(t) - \frac{2eV\tau_{r}}{\hbar}\cos\gamma(t)\right]$.


\section{Superconducting memristor}\label{sec:superconducting_memristor}

\subsection{Ideal case}\label{sec:ideal_memristor}

According to Eqs.~(\ref{eq:josephson2}) and~(\ref{eq:josephson1}), for an applied constant voltage bias, the phase-dependent dissipative current is oscillating and has zero average, and its amplitude is comparable to that of the Josephson current only at very high frequencies ($\sim 2\Delta/\hbar\sim 10^{11}\div 10^{12}\, {\rm Hz}$), therefore its detection is difficult.
The usual approach~\cite{pfl,vd,nw,rd,smp} has been the analysis of the damping of the plasma resonance~\cite{Dahm:1968} that, according to the RSJ model~(\ref{eq:josephson1}), has frequency $\omega_p(\gamma) = \sqrt{|\cos\gamma|/(L_cC)}$ [$L_c = \hbar/(2eI_c)$] and quality factor $Q(\gamma)=\omega_p(\gamma)C/[G_L(1+\varepsilon\cos\gamma)]$, the latter providing information on the PDC. Using the plasma resonance has the disadvantage that the resonance frequency itself is phase-dependent which can change the dissipative environment~\cite{Leppakangas:2011}. Moreover the PDC is an intrinsically nonlinear effect, while experiments  have probed the plasma resonance with a small AC current compared to the critical current --- well in the linear response regime.

We suggest a different way to access the PDC based on the use of a CA-SQUID as shown in Fig~\ref{fig1}\textbf{b}. The two junctions  of the interferometer (indexed by $i = 1,2$) have the same critical current $I_{c,1} = I_{c,2} = I_c$, but their electrodes have different values of the superconducting gap $\Delta_1 \neq \Delta_2$ and thus different normal conductances $G_{N,i}$ according to the Ambegaokar-Baratoff
result $G_{N,i}=2e I_c/(\pi\Delta_i)$~\cite{Ambegaokar:1963}. We define the ratio $r = G_L/G_N$ between the leakage conductance  (the conductance for $V<V_g$) and the normal conductance ($V>V_g$) which depends on the specifics of the junction. For standard ${\rm Nb}/{\rm AlO}_x/{\rm Nb}$ JJs one has $r\lesssim 0.1$, bur higher values can be attained by increasing the critical current density~\cite{Encyclopedia_Superconductors} (see also discussion in Sec.~\ref{sec:conclusions}). For simplicity we assume both the dimensionless quantities $\varepsilon$ and $r$ to be the same for the two junctions, a reasonable but not crucial assumption for what follows. Then the Ambegaokar-Baratoff relation implies that $G_{L,1}/G_{L,2} = \Delta_2/\Delta_1$.

The difference between the gauge invariant phases $\gamma_i$ of the two junctions is equal to the magnetic flux $\Phi$~\cite{Tinkham_book} through the loop
\begin{equation}
\gamma_1-\gamma_2= 2\pi \frac{\Phi}{\Phi_0}\,.
\end{equation}
If the loop inductance $L$ is small, namely $2\pi LI_c/\Phi_0 \ll 1$~\cite{Likharev_book}, the loop flux is equal to the external magnetic flux $\Phi = \Phi_{\rm ext}$. For $\Phi = \Phi_{\rm ext} = \Phi_0/2$ the Josephson currents in the two arms of the loop interfere destructively and cancel out. On the other hand the PDC is finite
\begin{gather}\label{eq:squid_memristor}
I = G_L'\left[1+ \varepsilon'\cos\gamma\right]V,\quad\quad\frac{d\gamma}{dt} = \frac{2\pi}{\Phi_0} 
\end{gather}
with $G_L' = G_{L,1}+G_{L,2}$, $\gamma = \gamma_1 = \gamma_2+\pi$ and
\begin{equation}\label{eq:epsilon_prime}
 \varepsilon' = \varepsilon\frac{G_{L,1}-G_{L,2}}{G_{L,1}+G_{L,2}}\,.
\end{equation}
We use the convention that primed quantities refer to the CA-SQUID considered as a single lumped JJ. The effect of capacitance and the fluctuations have been ignored in Eqs.~(\ref{eq:squid_memristor}) and will be addressed in the following. According to Eq.~(\ref{eq:epsilon_prime}) the PDC is necessarily zero, or very small, in an interferometer with JJs made of the same material.
Eqs.~(\ref{eq:squid_memristor}) have the same form of Eq.~(\ref{eq:memristor1}) that defines an ideal memristor~\cite{chua3,chua1,Pershin:2011}. Physically, the internal memory degree of freedom that controls the conductance is the non-dissipative current that flows in the loop, since opposite currents with equal magnitude flow in the two junctions. This loop current has no preferred values since the critical currents of the junctions are strictly equal.
On the other hand a critical current imbalance between the junctions introduces a potential term $-E_J\cos\gamma$ in the interferometer energy with $E_J = \hbar|I_{c,1}-I_{c,2}|/(2e)$ and the zero-current state $\gamma = 0$ is favoured.

\subsection{Realistic superconducting memristors}\label{sec:experimental_consideration}
The main assumptions used in the derivation of the defining equation~(\ref{eq:squid_memristor}) of a superconducting memristor are:
\begin{enumerate}
\item nearly equal critical currents of the two junctions $I_{c,1} \approx I_{c,2}$;
\item negligible capacitance, namely small Steward-McCumber parameter $\beta_c = 2eI_cC/(\hbar G_{N}^2) \lesssim 1$; 
\item small loop inductance $\lambda =2\pi LI_c/\Phi_0 \ll 1$;
\item external flux fixed at half flux quantum $\Phi_\text{ext}=\Phi_0/2$.
\end{enumerate}

The frequency window where the phase-dependent dissipative current (together with the quasiparticle current) dominates other current components is provided by 
conditions \textbf{1} and \textbf{2} and reads 
\begin{equation}\label{eq:window}
\frac{2e|I_{c,1}-I_{c,2}|}{|\varepsilon'|r \,\hbar G_N'} =\frac{i_c}{|\varepsilon'|r}\omega_c' < \omega < |\varepsilon'|r\frac{G_N'}{C'} = |\varepsilon'|r\frac{\omega_c'}{\beta_c'}\,,
\end{equation}
with $i_c = \left|\frac{I_{c,1}-I_{c,2}}{I_{c,1}+I_{c,2}}\right|$ the critical current suppression factor.
We used the total normal conductance $G_N' = G_{N,1} + G_{N,2}$ and total capacitance $C' = C_1 +C_2$ of the interferometer, while $\omega_c'=2e(I_{c,1}+I_{c,2})/(\hbar G_N')$ and $\beta_c' = 2e (I_{c,1}+I_{c,2})C'/(\hbar G_N'^2)$ are its characteristic frequency and Steward-McCumber parameter. For nearly identical critical currents $I_{c,1}\approx I_{c,2}$ one can easily relate the primed quantities to the corresponding (unprimed) ones for the two JJs, namely $\omega_c' = 2(\omega_{c,1}^{-1}+\omega_{c,2}^{-1})^{-1}$ and $\beta_c' = 2(\beta_{c,1}\Delta_2^2+\beta_{c,2}\Delta_1^2)/(\Delta_1+\Delta_2)^2$. In the limit $\Delta_1 \ll \Delta_2$ the above expressions reduce to $\omega_c' \sim 2\omega_{c,1}$ and $\beta_c' \sim 2\beta_{c,1}$. In fact, in this limit the properties of the whole device, including the PDC, are essentially those of the JJ with smaller superconducting gap, while the JJ with larger gap only serves as a shunt of the Josephson current $I_S$. Therefore, in order to construct a superconducting memristor, it is not necessary to achieve the same level of control on the parameters of both junctions, which is usually hard when different superconducting materials are employed.

The most important parameter is $i_c $ in Eq.~(\ref{eq:window}) since the lower the frequency the easier is to measure  directly the PDC. With current junction fabrication technology, $i_c$ is at best $\sim 10^{-2}$, while using a balanced SQUID~\cite{Kemppinen:2008} it is possible to obtain $i_c = 10^{-3}\div 10^{-4}$. The parameters $|\varepsilon'|$ and $r$ are already close to unity and can not be controlled easily. It follows that the lower frequency at which the PDC would manifest itself is $\omega \sim 10^{8}\div 10^{9}\, {\rm Hz}$ for values of $\omega_c\sim {\rm min}[\Delta_1,\Delta_2]/\hbar$ typical of low-$T_c$ superconductors, a substantial improvement. This sets also the scale for the voltage $V\sim 1 \mu{\rm V}$ since we will see below that the condition $2eV/{(\hbar \omega)}\sim 1$ must be satisfied in order to observe interesting effects. A further decrease of the frequency would lead to an impractical low value of the voltage.

It is not necessary to have a very low $\beta_c'$ unless the device is required to operate at high frequencies (see Fig.~\ref{fig:imperfect} below). It is important to notice that in our case $\beta_c'$ can not be lowered with a shunt resistance, since $|\varepsilon'|$ would be decreased as well. Low values of $\beta_c$ are routinely obtained with unshunted ${\rm Nb}/{\rm AlO}_x/{\rm Nb}$ junctions with critical current density $j_c \gtrsim 100\, {\rm kA/cm}^2$  that are employed in the fastest RSFQ circuits~\cite{Chen:1999,Hidaka:2006} (see also discussion in Sec.~\ref{sec:conclusions}).
Niobium and aluminum are a suitable choice for the superconducting material with smaller gap, since the fabrication technology for this kind of JJs is well developed.
Possible choices for the superconductor with larger gap are, e.g., NbN~\cite{Encyclopedia_Superconductors} and MgB$_2$~\cite{Xi:2008,Cunnane:2013}.

The condition \textbf{3} on the loop inductance is routinely realized in practical superconducting circuits, e.g. SQUID-based magnetometers~\cite{Likharev_book}. In order to enforce condition \textbf{4} on the flux bias, an alternative to an external magnetic field is to use superconductor-ferromagnet-superconductor junctions that induce a $\gamma = \pi$ phase drop across the electrodes in their zero-current state ($\pi-$junctions)~\cite{Ustinov:2003,Feofanov:2010}, with the advantage of reducing the loop inductance and the size of the device, and mitigating the effect of magnetic noise.

The effect of a small deviation from each of the above conditions will be considered in more detail in the following section.


\begin{figure}
\includegraphics{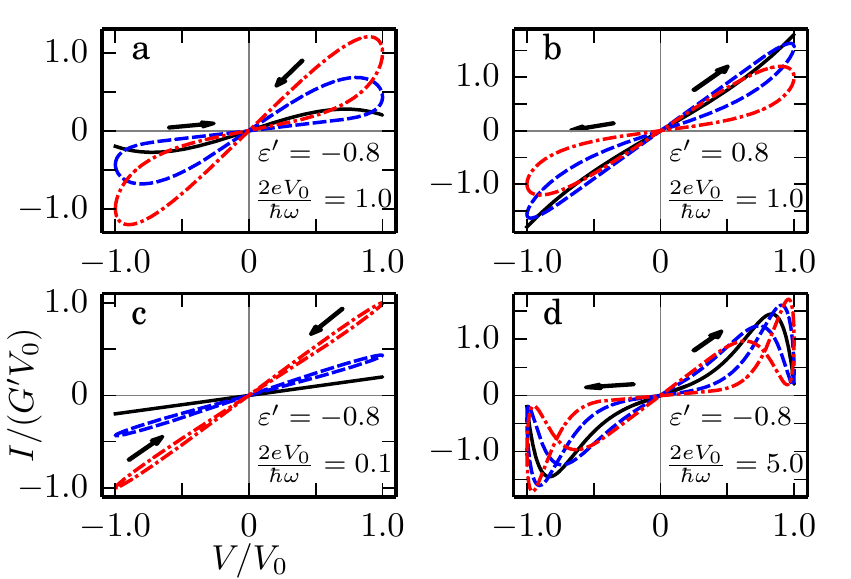}
\caption{\label{fig2} Pinched hysteresis loops in the $I-V$ plane for a superconducting memristor with phase of the form $\gamma(t) = \gamma_0 + \frac{2eV_0}{\hbar \omega}\sin\omega t$, voltage $V(t) = V_0\cos\omega t$ and current given by Eq.~(\ref{eq:pinched}). The solid line refers to $\gamma_0 =0$, dashed line $\gamma_0=\pi/4$ and dashed-dotted line $\gamma_0=\pi/2$. The hysteresis is well visible for $2eV_0/(\hbar \omega) = 1.0$ in panels (\textbf{a}) ($\varepsilon'=-0.8$) and (\textbf{b})  ($\varepsilon'=0.8)$. For $2eV_0/(\hbar \omega) =0.1$ [panel (\textbf{c})] the hysteresis is suppressed. Additional crossings appear upon increasing $2eV_0/(\hbar \omega)$ [panel (\textbf{d})].}
\end{figure}

\subsection{Pinched hysteresis loops}\label{sec:pinched_hysteresis_loop}
Our essential prediction is that in a CA-SQUID in a properly tuned magnetic field the gauge-invariant phase $\gamma$ affects the conductance even in the absence of a Josephson current. This is revealed by hysteresis loops  that pass through the origin of the $I-V$ plane under periodic driving. These so-called \textit{pinched hysteresis loops} are a fingerprint of memristive systems~\cite{chua3,chua2,chua1,Pershin:2011,myNanotech2013,strukov}. The zero-crossing property of hysteresis loops is equivalent to the property of zero energy storage in the circuit element~\cite{chua2}.  A pinched hysteresis loop is a nonlinear effect that can not be observed in experiments performed in the linear response regime, such as those that take advantage of the plasma resonance.

The current given by Eq.~(\ref{eq:squid_memristor}) when the phase has the form  $\gamma(t) = \gamma_0 + \frac{2eV_0}{\hbar \omega}\sin\omega t$ is
\begin{equation}\label{eq:pinched}
\frac{I(t)}{G_L'V_0} = \left[1+\varepsilon'\cos\left(\gamma_0+\frac{2eV_0}{\hbar \omega}\sin\omega t\right)\right]\cos\omega t\,,
\end{equation}
and it is shown against the voltage $V(t) = V_0\cos\omega t$ in Fig.~\ref{fig2} for different values of $\gamma_0$, $\varepsilon'$ and $2eV_0/(\hbar \omega)$. Several pinched hysteresis loops are visible for $2eV_0/(\hbar \omega)\gtrsim 1$. A unique property of superconducting memristors is that the conductance is a periodic and even function
\begin{equation}\label{eq:periodicity}
G(\gamma) = G(-\gamma) = G(\gamma+2\pi)\,.
\end{equation}
The definite parity implies that $\gamma_0$ must be different from $0$ or $\pi$ for the loop to enclose a finite area,  
while the periodicity manifests itself in the additional crossings of the two branches of the loop (see Fig.~\ref{fig2}\textbf{d}). The constant phase difference $\gamma_0$ is not associated to a finite Josephson current, but nevertheless it can be controlled externally as usual, i.e., by inserting the superconducting memristor into a superconducting loop whose threading flux is externally tuned. In the same way it is possible to produce a time-dependent phase $\gamma(t)$.

\begin{figure*}
\includegraphics[scale=1]{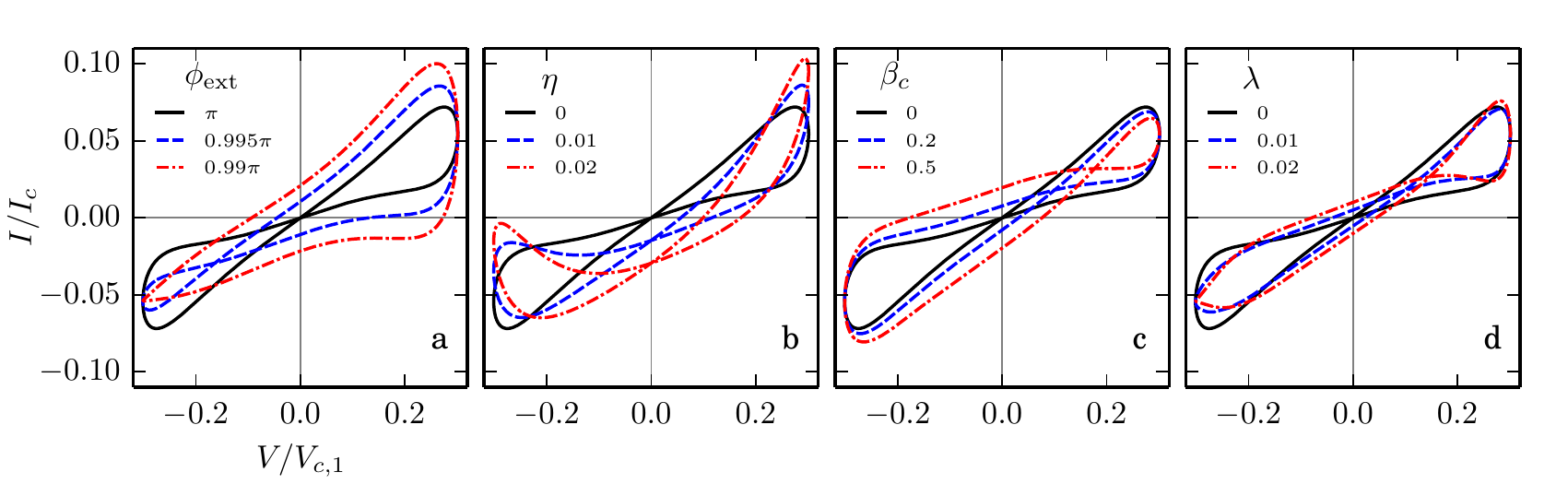}
\caption{\label{fig:imperfect}Impact on pinched hysteresis loops of deviations from the conditions discussed in Sec.~\ref{sec:experimental_consideration}. We considered in panel (\textbf{a}) a flux $\Phi_{\rm ext} =\Phi_0\frac{\phi_{\rm ext}}{2\pi}$ slightly different from $\Phi_0/2$; (\textbf{b}) a critical current imbalance $\left.I_{c}\right|_{1,2} = I_c(1\pm\eta)$; (\textbf{c}) nonzero Steward-McCumber parameters $\beta_{c,1}=\beta_{c,2}\neq 0$; (\textbf{d}) a finite loop inductance $\lambda = 2\pi LI_{c}/\Phi_0>0$ . The CA-SQUID is phase-biased as in Fig.~\ref{fig2} and we use the RSJ model~(\ref{eq:josephson1}) for the junctions. The voltage is measured in units of the characteristic frequency $V_{c,1} = I_{c,1}G_{N,1}^{-1}$, time in units of $\omega_{c,1}^{-1} = \hbar/(2eV_{c,1})$ and the current in units of $I_c$. The parameters used in the simulations are $r = 0.17$ [corresponding to $\kappa=0.1$ in Eqs.~(\ref{eq:Ip}) and~(\ref{eq:Iq})], $\omega/\omega_{c,1} = 1/50$,  $\gamma_0 = \pi/2$, $2eV_0/(\hbar\omega) = 2.4$, $\varepsilon=-0.56$, $G_{L,1}/G_{L,2} =\Delta_2/\Delta_1 = 5.0$, so according to Eq.~(\ref{eq:epsilon_prime}) $\varepsilon' = \frac{2}{3}\varepsilon = -0.37$. For small enough deviations  the crossing of the hysteresis loop moves away from the origin. The single-crossing property is however a relatively stable topological feature.}
\end{figure*}

The conditions discussed previously are necessary and sufficient for a superconducting memristor to show zero-crossing pinched hysteresis loops. In the four panels of Fig.~\ref{fig:imperfect}, we investigate  the effect of lifting each of the above requirements one at time. Quite generally we note that, if the deviations from the ideal values are small, the effect on the hysteresis loop is a shift of the crossing away from the origin. The property of \textit{single crossing} as a topological feature is remarkably robust and survives deviations from the dimensionless ideal values up to 0.01 in all cases. Such a tolerance is within reach of experiments. Moreover we observed that the property of a \textit{single crossing} or odd number of crossings in the $I-V$ plane can only be produced when the PDC dominates other contributions to the current and is a characteristic feature thereof. For example, by increasing the parameter $\lambda$ can lead to crossings that, however, always appear in pairs (an hint of this effect is visible in Fig.~\ref{fig:imperfect}\textbf{d} where a cusp appears in the curve for $\lambda = 0.02$).

\subsection{Non-destructive readout}\label{sec:nondestructive_readout}

A finite PDC is unambiguously revealed by pinched hysteresis loops under a periodic driving.
The periodicity property~(\ref{eq:periodicity}) allows for another way to detect the PDC using a single-flux-quantum voltage pulses.
Single-flux-quantum voltage pulses are defined by $\int dt\, V(t) = \Phi_0$ and in RSFQ logic are used to carry a unit of information (bit). When such a pulse is applied to a device described by Eq.~(\ref{eq:squid_memristor}) the total phase jump is asymptotically $2\pi$, which means that the internal state is unchanged.

\begin{figure}
\includegraphics{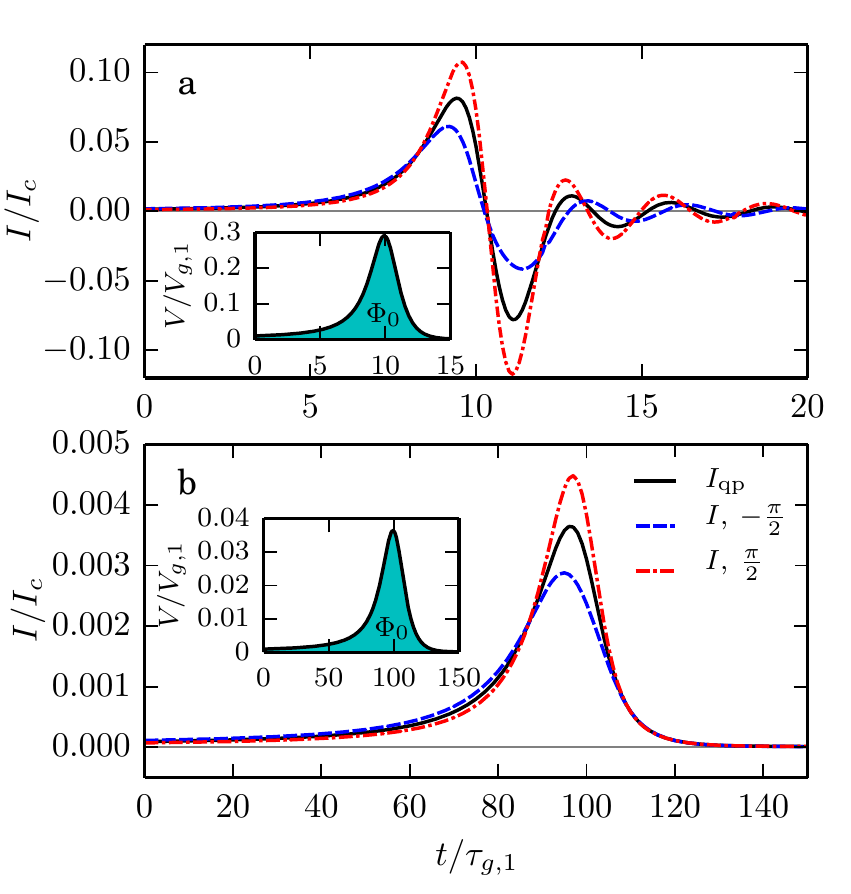}
\caption{\label{fig3} Current pulse through a superconducting memristor biased with a single-flux-quantum voltage pulse (shown in the insets, $V_{g,1} = 2\Delta_1/e$) calculated using the TJM model (the parameters used are $\Delta_2/\Delta_1 = G_{L,1}/G_{L,2} = 5$ and $\varepsilon' \approx -0.22$). Panel \textbf{a} refers to a fast pulse with width $\sim 3\tau_{g1}$ while the same pulse has been dilated in time by a factor of 8 in panel \textbf{b} while preserving the area $\int dt\, V=\Phi_0$. The solid line is the quasiparticle current $I_\text{qp}$ which is independent of the initial phase, while the dashed and the dash-dotted line are the total current  $I=I_\text{qp}+I_\text{pair}$ with initial state $\gamma(0) = -\pi/2$ and $\pi/2$, respectively.}
\end{figure}

However different initial values $\gamma_0$ of the phase affect the output current in a measurable way. This is demonstrated in Fig.~\ref{fig3} where we show the current through a superconducting memristor induced by two single-flux-quantum voltage pulses with time scales $\sim 3\tau_{g,1}$ and $\sim 25\tau_{g,1}$ respectively ($\tau_{g,1} = \hbar/\Delta_1$), obtained using the TJM model~\cite{Harris:1976,Likharev_book} since for time scales $\sim \tau_{g,1}$ the RSJ model~(\ref{eq:josephson1}) is inaccurate. While the quasiparticle component of the current $I_\text{qp}$ is independent of $\gamma_0$, the maximum  pair current $I_\text{pair}$ is different for the different initial states $\gamma_0 =-\pi/2$ (dashed line in Fig.~\ref{fig3}) and $\gamma_0 = \pi/2$ (dashed-dotted line). The current swing for different initial states $\gamma_0 \in [0,2\pi]$ in the limit of wide voltage pulses ($\gtrsim 20\tau_{g1}$ as in Fig.~\ref{fig3}\textbf{b}) is given by the parameter $\varepsilon'$ in Eq.~(\ref{eq:squid_memristor})---the microscopic model and the RSJ model give similar results in this case. In Fig.~\ref{fig3} we used $\varepsilon'\approx -0.22$, a small value compared to experiments~\cite{pfl,fpt,vd,nw,rd}.

We call this protocol \textit{non-destructive readout} since the phase $\gamma_0$ is measured  without changing it. This protocol represents a possible working principle for the read operation in a superconducting memory.

\subsection{Fluctuations}\label{sec:fluctuations}

We now explore the role of fluctuations on the behavior of a CA-SQUID. Although we reserve a thorough
description of fluctuations for future work, here we consider a model that captures their main features and confirms that the superconducting memristor we suggest is quite robust against fluctuations.

A description of a superconducting memristor that includes fluctuations and a finite capacitance $C'$ is provided by the system of Langevin equations
\begin{gather}
C'\frac{dV}{dt} + G_L'(1+\varepsilon'\cos\gamma)V + I_F = I\,,\label{eq:lang1}\\
\frac{d\gamma}{dt} = \frac{2e}{\hbar}V\,,\label{eq:lang2}
\end{gather}
with $\langle I_F(t) \rangle = 0$ and the noise autocorrelation function given by
\begin{equation}\label{eq:autocorrelation}
\left\langle I_F(t) I_F(t')\right\rangle = 2 k_\text{B}T G_L'(1+\varepsilon'\cos\gamma(t'))\delta(t-t')\,,
\end{equation}
which can be derived from the microscopic theory in the case of an arbitrary given phase dynamics $\gamma(t)$~\cite{Zorin:1981,Likharev_book}. Eq.~(\ref{eq:autocorrelation}) holds in the limit $|t-t'| \gg \tfrac{\hbar}{k_\text{B}T}\sim\frac{\hbar}{\Delta_1}$, which is appropriate in this case since we are interested in the low frequency dynamics of the superconducting memristor. Note that, in view of the dynamics of the superconducting phase, unlike Ref.~\cite{Rogovin:1974} where a \textit{time-averaged} autocorrelation function was considered, in the present work we use a non-averaged one. If a time average of Eq.~(\ref{eq:autocorrelation}) is taken the cosine term vanishes in the case of a DC voltage bias.

A Fokker-Planck equation for the phase space probability distribution $\sigma(\gamma,V,t)$ is then unambiguously associated to Eqs.~(\ref{eq:lang1})-(\ref{eq:lang2}), namely
\begin{equation}
\begin{split}
&\frac{\partial \sigma}{\partial t}
+ \frac{2e}{\hbar}\frac{\partial}{\partial \gamma}\left(\sigma V \right) + \frac{1}{C'}\frac{\partial}{\partial V}\left(\sigma I\right) \\ &= \frac{G_L'}{C'}(1+\varepsilon'\cos\gamma)\frac{\partial}{\partial V}\left[\left(V+\frac{k_\mathrm{B}T}{C'}\frac{\partial}{\partial V}\right)\sigma\right]\,.\label{eq:fokker-planck}
\end{split}
\end{equation}
We say unambiguously since the fluctuation term is nonlinear, i.e., it depends on the phase $\gamma(t)$, which usually creates an ambiguity in the interpretation of a stochastic equation (the It\^o-Stratonovich dilemma), but this is not the case here. In fact, a system of first order Langevin equations in the variables $y_{\nu}$ with fluctuation term $C_{\nu}(y)L(t)$ ($L(t)$ is white noise) is free of ambiguity if the condition $\sum_{\mu}C_\mu(y)\frac{C_\nu(y)}{\partial y_\mu} = 0$ is satisfied and then a unique Fokker-Planck equation is associated with it~\cite{vanKampen_book}. The previous condition is satisfied for Eqs.~(\ref{eq:lang1}) and~(\ref{eq:lang2}) only with a finite capacitance $C'$.

It would be desirable to have an approximate equation for the probability distribution of $\gamma$ alone, namely $\sigma(\gamma,t) = \int dV\, \sigma(\gamma,V,t)$ since we are interested in the limit of overdamped junctions ($C'\to 0$). Taking this limit in the Langevin system~(\ref{eq:lang1})-(\ref{eq:lang2}) leads to the problem mentioned above. A more controlled procedure is to take the limit of small capacitor discharge time scale $\tau_{RC} =  C'/G_L'$ in the Fokker-Planck equation~(\ref{eq:fokker-planck}). This can be calculated using  the technique for the elimination of fast (with time scale $\tau_{RC}$) variables  of Ref.~\cite{vanKampen:1985}. The starting point for the method detailed in Ref.~\cite{vanKampen:1985} is to write Eq.~(\ref{eq:fokker-planck}) in the general form
\begin{equation}
\frac{\partial \sigma}{\partial t} = \left(\frac{1}{\tau_{RC}}\mathcal{L}^{(0)} + \mathcal{L}^{(1)}\right)\sigma\,,
\end{equation}
where the differential operators $\mathcal{L}^{(i)}$ are defined by
\begin{gather}
\mathcal{L}^{(0)} = (1+\varepsilon'\cos\gamma)\left(\frac{\partial}{\partial V}V +\frac{k_\mathrm{B}T}{C'}\frac{\partial^2} {\partial V^2}\right)\,,\\
 \mathcal{L}^{(1)} = -\frac{2eV}{\hbar}\frac{\partial}{\partial \gamma}
- \frac{I}{C'}\frac{\partial}{\partial V}\,.
\end{gather}
The case $\varepsilon' = 0$ is worked out step by step in Ref.~\cite{vanKampen:1985} and the same procedure can be applied for finite $\varepsilon'$ since the prefactor $(1+\varepsilon'\cos\gamma)$ commutes with the partial derivative $\partial/\partial V$. The final result at first order in $\tau_{RC}$ is a drift-diffusion equation for the phase
\begin{equation}
\label{eq:diffusion}
\begin{split}
&\frac{\partial \sigma}{\partial t}=  \frac{2e}{\hbar G'_L}\frac{\partial}{\partial \gamma}\left[ \frac{1}{1+\varepsilon' \cos\gamma}\left(-I+\frac{2ek_{\mathrm{B}}T}{\hbar}\frac{\partial}{\partial \gamma}\right)\sigma\right].
\end{split}
\end{equation}

The diffusion coefficient [$\propto (1+\varepsilon'\cos\gamma)^{-1}$] is phase-dependent  for finite $\varepsilon'$ and the diffusion time scale is $\tau_{\rm Diff} = \hbar^2G_L'/(4e^2k_BT) = 7.84\,\mathrm{ns}/(R[\Omega]T[\mathrm{K}])$ with $R = 1/G_L'$. For large values of the total subgap conductance $G'_L$ (large junction critical current) and small enough temperature, $\tau_{\rm Diff}$ is more than three orders of magnitude larger than the typical time scale (picoseconds) of the phenomena discussed above. This justifies having neglected the effect of noise previously.
Higher orders corrections of Eq.~(\ref{eq:diffusion}) can be computed if necessary. The first order is a good approximation if $\tau_{\mathrm{RC}}/\tau_\mathrm{Diff} = 4e^2k_{\mathrm{B}}TC'/(\hbar^2G_L'^2)\ll 1$.

If the critical currents of the two junctions in Fig.~\ref{fig1}\textbf{b} are strictly equal, the phase $\gamma$ can take any value at no energy cost. Then thermal fluctuations  [$I_{F}(t)$ in Eq.~(\ref{eq:josephson1})] induce the brownian motion of the phase leading to the so-called \qql stochastic catastrophe\qqr~\cite{myNanotech2013}, i.e., the value of $\gamma$ is totally randomized. This process is described by Eq.~(\ref{eq:diffusion}) which is physically sound since at equilibrium the probability distribution for the phase is uniform, as expected for zero total critical current of the CA-SQUID. 

\section{Discussion and conclusion}\label{sec:conclusions}
We have shown how the PDC in JJs can be probed directly using a simple two-junction interferometer realized using a combination of different superconducting materials (CA-SQUID). The conditions for realizing our proposal do not rely on specific models since they are expressed in terms of general properties of superconducting weak links (see Sec.~\ref{sec:experimental_consideration}) and a good approximation thereof should be within reach of present JJ fabrication technologies. This allows the study of new manifestations of the PDC such as pinched hysteresis loops and the non-destructive readout of the initial phase with a single-flux-quantum voltage pulse. The equations that govern the dynamics of a CA-SQUID with vanishing critical current [Eq.~(\ref{eq:squid_memristor})] are an instance of the general memristor equations [Eqs.~(\ref{eq:memristor1})].

As we argue in Sec.~\ref{sec:pdc}, the PDC is a general phenomenon present in any kind of superconducting weak link, a possible reason being that the response of the supercurrent to a variation of the phase is not instantaneous but retarded, producing a phase-dependent dissipation. Indeed, experiments on non-tunnel junctions~\cite{vd,rd,nw,fpt,Likharev:1979} show that the ratio $\varepsilon$ between PDC and leakage conductance is almost unity, the largest allowed value. Therefore it should be possible to realize a superconducting memristor with junctions of the non-tunnel type which usually have higher transparency than tunnel junctions.
   
Increasing the barrier transparencies of JJs is expected to improve the performance of a superconducting memristor for several reasons: first, the PDC is of the same order of the leakage conductance which is increased by increasing the transparency. According to Eq.~(\ref{eq:window}) the window of frequencies where the PDC dominates is widened with increasing the ratio $r = G_L/G_N$. Second, the critical current density increases as well with the transparency and this is desirable since the non-dissipative current flowing in the CA-SQUID loop is, in our proposal, the physical memory degree of freedom of the memristor, and the larger the current the less sensitive is the state of the memristor to thermal fluctuations. Indeed the phase diffusion time scale defined in Sec.~\ref{sec:fluctuations} is proportional to the critical current of the junctions $\tau_\text{Diff} \propto I_c$. Third, the capacitance of junctions with high transparency is generally small and they are over-damped even without a shunting resistor.

This is the same kind of trend that has occurred with RSFQ technology. In this case, due to the development of the fabrication technology of Nb/AlO/Nb JJs used in RSFQ circuits, the barrier layer is so thin that transport is already  dominated by multiple Andreev reflections~\cite{Bunyk:2001}, a regime of high transparency not captured by the TJM model. The problem of the existence of a finite PDC for JJs where transport is dominated by Andreev reflections is an interesting question by itself. If a sizable PDC is present the use of this kind of junctions could improve the specifics of superconducting memristors.

A possible problem though is that in non-tunnel junctions the current-phase relation can be different from sinusoidal, which is the case for JJs that are well described by second-order perturbation theory on the tunneling matrix elements, and this results in an imperfect cancellation of the total critical current of the interferometer. Moreover, a general dynamical theory as informative as the TJM model~(\ref{eq:tjm1})-(\ref{eq:tjm2}) is not available for non-tunnel junctions.

Our proposal, besides being interesting from a fundamental point of view, may find practical applications as well. It has been found recently that unconventional nanoscale devices that combine electrical and ionic transport are approximately governed by the defining equations of a memristor~[Eqs.~(\ref{eq:memristor1})]~\cite{strukov,Pershin:2011}. This has stimulated the exploration of neuromorphic massively-parallel computing architectures~\cite{DiVentra:2013,Likharev:2011}, whose speed and low-energy consumption for specific tasks promise to be unmatched even by the best available computers. Superconducting memristors offer new venues for neuromorphic computation~\cite{DiVentra:2013,Chiarello:2012} and high-speed digital electronics since they can be readily integrated with existing RSFQ circuits with clock frequencies up to hundreds of Gigahertz~\cite{Chen:1999}, combined with $10^5$ times lower power consumption than their semiconductor counterparts. The non-destructive readout protocol presented in Sec.~\ref{sec:nondestructive_readout} may find applications in superconducting memories, a subject which is becoming increasingly actual~\cite{Heim:2013}.

\acknowledgements{This work has been supported by DOE under Grant No. DE-FG02-05ER46204 and the Center for Magnetic Recording Research at UCSD. We thank F. Giazotto for useful discussions, S. Gasparinetti, F. Mireles and K. K. Likharev for the critical reading of our paper.}

\appendix

\section{Sign and magnitude of the PDC within the exponential regularization}\label{appendix:analytic_pdc}

The integrals in Eq.~(\ref{eq:A_nm}) can be evaluated in closed form using the complete elliptic integrals of the first and second kind, respectively
\begin{gather}
K(z) = \int_0^{\frac{\pi}{2}} \frac{d\theta}{\sqrt{1-z\sin^2\theta}}\,,\\
E(z) = \int_0^{\frac{\pi}{2}} d\theta\,\sqrt{1-z\sin^2\theta}\,.
\end{gather}
We provide only the first few that are relevant for our problem:
\begin{align}
A_{0,0}(\kappa) &= -\frac{K\left(-\kappa ^2\right)}{\pi}\,,\\
A_{0,1}(\kappa) &= \frac{E\left(-\kappa ^2\right)-\left(1+\kappa ^2\right) K\left(-\kappa
^2\right)}{2 \pi  \kappa  \left(1+\kappa ^2\right)}\,,\\
A_{1,0}(\kappa) &= \frac{2 \kappa -2 E\left(-\kappa ^2\right)+K\left(-\kappa ^2\right)}{\pi
}\,,\\
A_{1,1}(\kappa) &= \frac{\left(1+2 \kappa ^2\right) E\left(-\kappa
^2\right)-\left(1+\kappa ^2\right) \left(2 \kappa +K\left(-\kappa ^2\right)\right)}{2 \pi  \kappa  \left(1+\kappa ^2\right)}\,.
\end{align}

\begin{figure}
\includegraphics{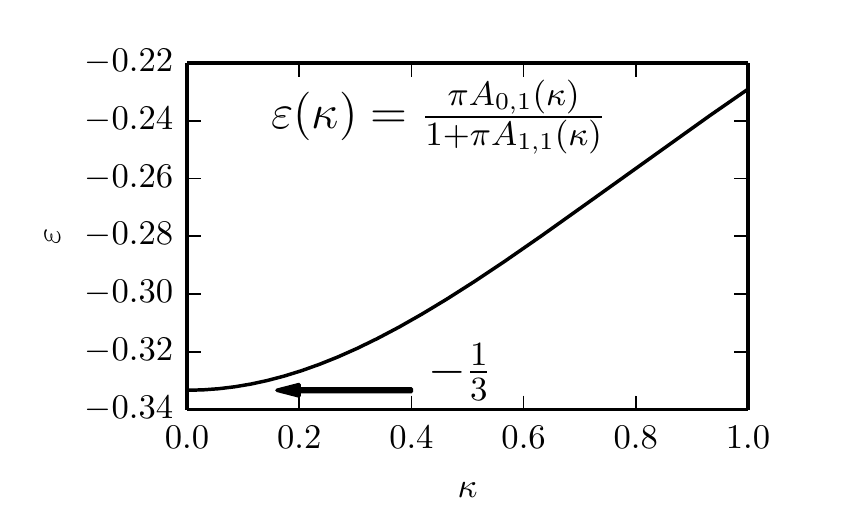}
\caption{\label{fig:epsilon} The function $\varepsilon(\kappa)$ as a function of the parameter $\kappa$. The arrow shows the limit $\lim_{\kappa\to 0}\varepsilon(\kappa) = -1/3$.}
\end{figure}

These formulas are useful to evaluate the ratio $\varepsilon(\kappa)= \pi A_{0,1}(\kappa)/(1+\pi A_{1,1}(\kappa))$ between the PDC and the leakage conductance [see Eq.~(\ref{eq:josephson1})]. Using that $K(z) \approx \frac{\pi}{2}\left(1+\frac{z}{4}\right)$ and $E(z) \approx \frac{\pi}{2}\left(1-\frac{z}{4}\right)$ for $z\to 0$, one finds that  $\lim_{\kappa\to 0}\varepsilon(\kappa) = -1/3$. This result was derived with a frequency-domain approach in Ref.~\cite{Zorin:1979}. The plot in Fig.~\ref{fig:epsilon} shows in fact that the function $\varepsilon(\kappa)$ does not change significantly in the relevant range $0\leq \kappa \leq 1$. Since the functions $J_{n}(t)Y_n(t)$ have the asymptotic behavior $\sim t^{-1}\cos 2t$ the integrals $A_{n,1}(0)$ are ill defined. This means that the magnitude and sign of the PDC strongly depend on the kind of regularization used for the BCS result, which is given by Eqs.~(\ref{eq:Ip})-(\ref{eq:Iq}) without the exponential factor. For example, with a different regularization used in Ref.~\cite{Zorin:1979} the result is $\varepsilon \approx -1$. Therefore  microscopic details, that provide such regularization and that are difficult to account for quantitavely, can have a dramatic effect on the PDC.

%

%
\end{document}